\documentclass[aps,prl,reprint,superscriptaddress]{revtex4-1}
\usepackage{amsmath,amssymb,graphicx,epspdfconversion,hyperref,cleveref}
\usepackage{natbib}
\usepackage[svgnames]{xcolor}
\hypersetup{hidelinks,colorlinks=true,allcolors=DarkBlue,pdfpagemode=UseNone}

\begin{document}

\title{Controlling interactions between quantum emitters using atom arrays}

\author{Taylor L. Patti}
\email[]{taylorpatti@g.harvard.edu}
\affiliation{Department of Physics, Harvard University, Cambridge, Massachusetts 02138, USA}

\author{Dominik S. Wild}
\affiliation{Department of Physics, Harvard University, Cambridge, Massachusetts 02138, USA}
\author{Ephraim Shahmoon}
\affiliation{Department of Physics, Harvard University, Cambridge, Massachusetts 02138, USA}
\affiliation{Department of Chemical \& Biological Physics, Weizmann Institute of Science, Rehovot 761001, Israel}

\author{Mikhail D. Lukin}
\affiliation{Department of Physics, Harvard University, Cambridge, Massachusetts 02138, USA}

\author{Susanne F. Yelin}
\affiliation{Department of Physics, Harvard University, Cambridge, Massachusetts 02138, USA}
\affiliation{Department of Physics, University of Connecticut, Storrs, Connecticut 06269, USA}

\begin{abstract}

We investigate two-dimensional atomic arrays as a platform to  modify the electromagnetic environment of individual  quantum emitters. Specifically, we demonstrate that control over emission linewidths,   
resonant frequency shifts, and local enhancement of driving fields is possible due to strong dipole-dipole interactions within ordered, subwavelength atom configurations. We demonstrate  that these effects can be used to dramatically enhance coherent dipole-dipole interactions between distant quantum emitters within an atom array. Possible 
experimental realizations and potential applications are discussed.  

\end{abstract}
\maketitle


High-fidelity, deterministic interactions between individual atoms and photons, as well as
photon-mediated interactions between quantum emitters, are central 
to many areas of quantum science and engineering  
\citep{Kimble2008, Chang2014, Benson2017, Hammerer2010, Pezze2018,Ma2011}. In free space, these interactions are limited by the scattering cross section of the emitter, which is typically bounded by a small geometrical limit \cite{Novotny2006}. To circumvent these  limits, optical cavities and waveguides can be utilized to enhance interaction 
probabilities \cite{Vetsch2010, Thompson2013, Gouraud2015, Goban2012, Reiserer2015}. Recent research has shown that photonic crystals can be used to efficiently engineer atom-photon interactions \cite{Gonzalez-Tudela2015, Douglas2015, Shahmoon2013, Calajo2016}. While substantial experimental progress towards the realization of these ideas has recently been made \cite{Hood2016, Liu2017, Sundaresan2019, Goban2014, Tiecke2014, Bhaskar2020}, widespread applications of these techniques remain limited by multiple obstacles. For instance, it is experimentally difficult to control cold, trapped atoms in the immediate proximity of nano-structured surfaces. Moreover, creating
and controlling a large number of identical solid-state quantum emitters within nanostructures constitutes an open experimental challenge.    


In this Letter, we demonstrate that two-dimensional arrays composed of localized emitters can be used to effectively engineer atom-photon interactions and to enable 
high-fidelity, long-range interactions between quantum emitters.  
 The resulting effective photonic materials would furnish inherent advantages, such as dynamic reconfigurability \cite{Anderlini2006},  large coherent coupling strength \cite{SuppInfo}, and an environment devoid of surface imperfections \cite{Chang2018}. Furthermore, these systems feature intrinsic quantum nonlinearities, which may enable intriguing fermion-like behavior of interacting photons \cite{Asenjo-Garcia2017, Zhang2019}.   While 1D atomic chains have been studied previously for coupling multiple target atoms of broad and delocalized linewidth \cite{Masson2019}, and similar investigations have focused on other quantum emitters, such as superconducting qubits \cite{Mirhosseini2019, Albrecht2019, Facchi2019}, only recently has it been shown that 2D atom arrays can interact strongly with individual photons \cite{Shahmoon2017, Bettles2016, Rui2019}. In particular, 2D arrays constitute promising candidates for a number of quantum information applications \cite{Shahmoon2017, Grankin2018, Zoubi2012, Asenjo-Garcia2017, Perczel2017, Bettles2016-2, Shahmoon2018}. \crefformat{equation}{Eq.~#2(#1)#3} 

\begin{figure}	
\includegraphics{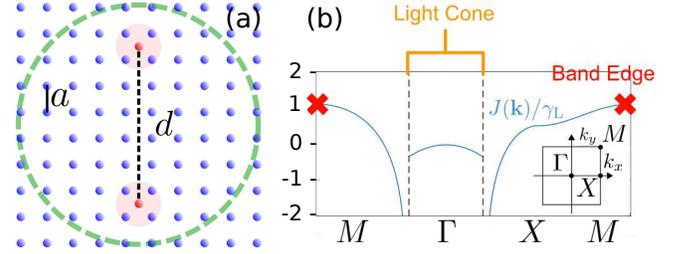}
\caption{\label{fig.Diagram} (a) 2D array of atoms (blue) of interatomic spacing $a \lesssim \lambda$, for lattice atom transition wavelength $\lambda$, with impurity atoms (red) embedded in-plane at plaquette centers and separated by distance $d$ (dashed line). While a free-space impurity (atomic qubit) has cross-section-limited light coupling (pink shading), its dipole-dipole interactions with the array extend over many lattice sites (dashed green circle). (b) Band structure $J(\mathbf{k})$ (see \cref{eq.LatticeCoupling}) of 2D atomic square lattice with spacing $a=0.2\lambda$.}
\end{figure}

The key idea of the present work can be understood by considering an individual 
impurity atom coupled to the 2D emitter array via 
dipole-dipole interactions. 
In the single excitation limit, these dipole-dipole interactions form normal modes on the 2D lattice via photon exchange. These modes  couple to the impurity, modifying its natural frequency $\omega_\text{I}$, linewidth $\gamma_\text{I}$, and local driving field Rabi frequency $\Omega_\text{I}$. In particular, this allows one to effectively confine 
and guide atomic emission within the 2D surface, and to create atom-photon bound states that can be used to engineer strong, coherent, and controllable interaction between 
distant impurities within the array. 


\crefformat{figure}{Fig.~#2#1{(b)}#3}

\begin{figure}
\begin{center}
\includegraphics[width=	\columnwidth]{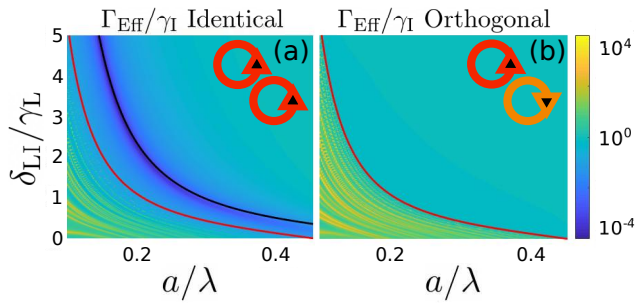}
\caption{(a) Renormalized impurity linewidth $\Gamma_{\text{Eff}}$ for identical and (b) orthogonal configurations (insets display relative impurity-array atom polarization) as a function of lattice spacing $a$ and array atom detuning $\delta_\text{LI}$. The plots shown are for a $20\times20$ array, however the impurity-array dynamics are nearly identical to those of an infinite lattice (see SM \cite{SuppInfo}). The band edge energy $\omega_\text{BE}$ is plotted in red. In (a), the black curve represents optimal lattice detuning $\delta^\text{D}_\text{LI}$ for suppressed impurity emission due to the lattice $\mathbf{k}=0$ mode. Enhanced emission primarily occurs in (a) and (b) for $\delta_\text{LI} < \omega_\text{BE}$ due to resonant coupling between the impurity and lattice modes.}
\label{fig.Gamma}
\end{center}
\end{figure}

We consider a system in which  the impurity atom is placed interstitially \cite{SuppInfo} within a square 2D atom array of spacing $a$, with $a \lesssim \lambda$, where $\omega_\text{L} = 2 \pi c / \lambda$ is the natural frequency of the lattice atoms (\crefformat{figure}{Fig.~#2#1{(a)}#3}\cref{fig.Diagram}).  Such subwavelength spacing is obtainable,  e.g., using ultracold atoms in an optical lattice \cite{Rui2019, Bloch2005,Labuhn2016}. Provided that $|\omega_\text{I} - \omega_\text{L}| \ll \omega_\text{I}$, $\omega_\text{L}$, the response of the atoms is approximated as a narrow peak around $\omega_\text{L}$ and we can employ the Green's tensor methods in Refs.\ \cite{Gruner1996, Dung2002, Buhmann2007, Buhmann2012}. We define the coherent $J(\mathbf{r}_i, \mathbf{r}_j)$ and dissipative $\Gamma(\mathbf{r}_i, \mathbf{r}_j)$ dipole-dipole interactions between \textit{any} two atoms $i$ and $j$ as the projection of the real and imaginary parts of the free-space Green's function onto the dipole transition unit vector $\hat{d}$ \cite{SuppInfo}. Excitations on an infinite lattice are collective surface modes of in-plane quasimomentum $\mathbf{k}$ with lowering operator $\sigma_\mathbf{k}$ and frequency shift and decay rate described by

\begin{equation}
J(\mathbf{k}) - \frac{i}{2} \Gamma(\mathbf{k}) = -\frac{3 \pi \gamma_\text{L} }{\omega_\text{L}}\hat{d^{\dagger}} \cdot\mathbf{G}(\mathbf{k},\omega_\text{L}) \cdot \hat{d},
\label{eq.LatticeCoupling}
\end{equation}

\noindent where $\mathbf{G}(\mathbf{k},\omega_\text{L})$ is the discrete Fourier Transformation of $\mathbf{G}(\mathbf{r}_i, \omega_\text{L})$ over the lattice sites. This formalism represents the lattice as a band structure of momentum modes (\crefformat{figure}{Fig.~#2#1{(b)}#3}\cref{fig.Diagram}), akin to those of photonic crystals. We indicate the highest energy level or band edge $\omega_\text{BE}$ as red crosses in \crefformat{figure}{Fig.~#2#1{(b)}#3}\cref{fig.Diagram} and red curves in \crefformat{figure}{Fig.~#2#1{(a)}#3}\cref{fig.Gamma} and (b), and the lower momentum modes that couple to far-field light or light cone as the yellow region of \crefformat{figure}{Fig.~#2#1{(b)}#3}\cref{fig.Diagram}, which is defined by $\mathbf{k}^2 \leq \omega_\text{L}^2$ for $c=1$. For lattice spacing $a < \lambda/\sqrt{2}$, there exist guided modes that cannot decay and propagate along the lattice without loss \cite{Shahmoon2017, Asenjo-Garcia2017}. The coupling of the impurity to any surface mode $\mathbf{k}$ is described by the analogous terms $\tilde{J}(\mathbf{k})$ and $\tilde{\Gamma}(\mathbf{k})$, as derived in the Supplementary Material (SM) \cite{SuppInfo}. The non-Hermitian Hamiltonian of the lattice with an impurity of lowering operator $s$ under the rotating wave approximation is

\begin{multline}
H = -i\frac{\gamma_\text{I}}{2}s^\dagger s - \sum_\mathbf{k} \left[\delta_\text{LI} \text{ - } J(\mathbf{k}) \text{+} i\frac{\Gamma(\mathbf{k})}{2}\right]\sigma^\dagger_\mathbf{k}\sigma_\mathbf{k} \\
+ \sum_\mathbf{k}\left[\tilde{J}(\mathbf{k}) \text{ - } i\frac{\tilde{\Gamma}(\mathbf{k})}{2}\right]\sigma^\dagger_\mathbf{k}s + \sum_\mathbf{k}\left[\tilde{J}(\mathbf{k})^* \text{ - } i\frac{\tilde{\Gamma}(\mathbf{k})^*}{2}\right]s^\dagger\sigma_\mathbf{k},
\label{eq.H}
\end{multline}

\noindent where $\delta_\text{LI} = \omega_\text{I}-\omega_\text{L}$. \crefformat{equation}{Eq.~#2(#1)#3}\cref{eq.H} holds as long as the retardation of light within the spatial scale of our system is negligible \cite{Chang2012}.

System dynamics are sensitive to the relative polarization of the lattice and impurity atoms, which determines the strength of $\tilde{J}(\mathbf{k})$ and $\tilde{\Gamma}(\mathbf{k})$ for each mode. We assume that all atoms have either right or left-handed circular polarization in the $xy$-plane. Furthermore, we identify the two polarization configurations that are key to this work: (1) the identical configuration, where both the lattice and impurity atoms have the same polarization (e.g. both right-handed) and (2) the orthogonal configuration, where the lattice and impurity atoms have the opposite polarization (e.g. right and left-handed, respectively, see Fig. S2). These polarizations could be individually addressed by inducing Zeeman shifts with a $z$-axis magnetic field. We emphasize that the orthogonal configuration still leads to impurity-lattice interaction, as these polarizations are only orthogonal along the $z$-axis, not within the $xy$-plane.

In order to gain physical intuition for the distinct effects of these two polarization configurations, we study a toy model: an impurity in a $2\times2$ atom array. This model is derived in the SM \cite{SuppInfo}. The impurity only couples to two of four array modes, specifically, $\hat{v}_\parallel$, the lowest momentum mode with the largest radiative linewidth, and $\hat{v}_\perp$, the highest momentum mode with the narrowest linewidth. In $\hat{v}_\parallel$, all atoms oscillate in-phase, whereas in $\hat{v}_\perp$ they oscillate $\pi$ out-of-phase in a checkerboard pattern. These modes form the characteristic points of a band structure (\crefformat{figure}{Fig.~#2#1{(b)}#3}\cref{fig.Diagram}), with $\hat{v}_\parallel$ at the center of the light cone and $\hat{v}_\perp$ at the band edge. An impurity in the orthogonal configuration only couples to $\hat{v}_\perp$ and an impurity in the identical configuration only couples to $\hat{v}_\parallel$. In this latter combination, the impurity and array oscillate $\pi$ out-of-phase from one another. This state is qualitatively comparable to a dark state in V-type electromagnetically induced transparency \cite{Fleischhauer2005}. The orthogonal configuration impurity couples to $\hat{v}_\perp$, forming a bright state. While a few-atom system may seem too simplistic to model the dynamics of atomic lattices, the modification of the impurity's electromagnetic environment converges with relatively few atoms, with $6\times6$ arrays producing electromagnetic effects that are nearly indistinguishable from those of arrays orders of magnitude larger \cite{SuppInfo}. 

Provided that $\gamma_\text{I} \ll \gamma_\text{L}$, the array's dynamics occur on a time scale much shorter than that of the impurity, rendering it a Markovian bath. The impurity will exchange photons, both real and virtual, with the array, giving rise to so-called self-interaction terms through which the impurity is influenced by the effect of its own presence in the array. In the weak driving limit, we approximate the lattice atoms as harmonic oscillators in steady state that produce the self-energy

\begin{equation}
\Sigma_{\text{SE}} = \sum_\mathbf{k} \frac{\left(\tilde{J}(\mathbf{k})-\frac{i}{2}\tilde{\Gamma}(\mathbf{k})\right)\left(\tilde{J}(\mathbf{k})^*-\frac{i}{2}\tilde{\Gamma}(\mathbf{k})^*\right)}{\delta_\text{LI} - J(\mathbf{k}) + \frac{i}{2}\Gamma(\mathbf{k})}.
\label{eq.SE}
\end{equation}

\noindent The self-energy is key to understanding impurity-lattice interactions as it modifies the effective frequency and decay rate of the impurity to $\omega_\text{Eff} = \omega_\text{I} + \mathbf{Re}[\Sigma_{\text{SE}}]$ and $\Gamma_\text{Eff} =\gamma_\text{I} - 2\mathbf{Im}[\Sigma_{\text{SE}}]$, respectively. These equations are valid as long as $\Sigma_\text{SE}$ varies little on the interval $\delta_\text{LI} + \mathbf{Re}[\Sigma_{\text{SE}}] \pm \Gamma_\text{Eff}$, such that the electromagnetic response of the lattice atoms with respect to $\delta_\text{LI}$ is approximately constant compared to that of the impurity atom. Under these same conditions, $\omega_\text{Eff} - \omega_\text{L} \approx \delta_\text{LI}$. We note that $\Sigma_\text{SE}$ can vary considerably over broad $\Gamma_\text{Eff}$, such as that of the orthogonal configuration near the band edge, and in such cases non-Markovian analysis may be valuable \cite{Kofman1994, Shahmoon2013}.

\crefformat{figure}{Fig.~#2#1{(a)}#3}\cref{fig.Gamma} displays $\Gamma_\text{Eff}$ in the identical configuration. Below $\omega_\text{BE}$ (red curve), $\Gamma_\text{Eff}$ is enhanced as the impurity couples to resonant lattice modes, particularly those in the light cone. Above $\omega_\text{BE}$, however, the linewidth of these states is suppressed by orders of magnitude due to destructive interference between the radiation of the impurity and off-resonant coupling with these modes. We can maximize the impurity lifetime (creating a ``dark" state as explained above) due to a particular mode with momentum $\mathbf{k}$ by minimizing the corresponding term in $\Gamma_\text{Eff}$ with respect to $\delta_\text{LI}$. As we place impurities at a plaquette center, $\tilde{J}(\mathbf{k})$, $\tilde{\Gamma}(\mathbf{k})$ are real, and we obtain optimized lattice detuning

\begin{equation}
\delta^\text{D}_\text{LI}(\mathbf{k}) = J(\mathbf{k}) - \frac{\tilde{J}(\mathbf{k})\Gamma(\mathbf{k})}{\tilde{\Gamma}(\mathbf{k})}.
\label{eq.deltaoptimal}
\end{equation}

\noindent This quantity is plotted in black in \crefformat{figure}{Fig.~#2#1{(a)}#3}\cref{fig.Gamma} for $\mathbf{k}=0$ and corresponds to the curve of smallest $\Gamma_\text{Eff}$ and largest excitation probability. The correspondence of $\mathbf{k}=0$ demonstrates that light cone coupling dominates identical configuration dynamics. In the SM, we show that linewidth suppression is lattice spacing limited, as $\Gamma_\text{Eff} \rightarrow 0$ in the limit $a/\lambda \ll 1$, while $\delta^\text{D}_\text{LI} \propto 1/a^3$ \cite{SuppInfo}.

\crefformat{figure}{Fig.~#2#1{(b)}#3}\cref{fig.Gamma} depicts $\Gamma_\text{Eff}$ in the orthogonal configuration. Like in the identical configuration, $\Gamma_\text{Eff}$ is enhanced due to impurity coupling to resonantly driven lattice modes for $\delta_\text{LI} < \omega_\text{BE}$. In the orthogonal configuration, however, $\Gamma_\text{Eff}$ enhancement is greater and occurs near resonance with band edge states, rather than those of the light cone. 

In the presence of an incident driving field with lattice mode Rabi frequency $\Omega_\text{L}(\mathbf{k})$ such that $\Omega_\text{L}(\mathbf{k})/\gamma_\text{L} \ll 1$, the impurity is influenced by the driving of the modes to which it couples and thus experiences a lattice-mediated field with effective Rabi-frequency

\begin{equation}
\Omega_\text{Eff} = \sum_\mathbf{k} \frac{\left(\tilde{J}(\mathbf{k})+\frac{i}{2}\tilde{\Gamma}(\mathbf{k})\right)\Omega_\text{L}(\mathbf{k})}{\delta_\text{LI} - J(\mathbf{k}) - \frac{i}{2}\Gamma(\mathbf{k})} + \Omega_\text{I},
\label{eq.LMR}
\end{equation}

\noindent where $\Omega_\text{I}$ is the Rabi drive of the impurity in free space and we assume the drive to be resonant with the impurity. This lattice mediation can be destructive, isolating the impurity from the incident drive such that $\Omega_\text{Eff} \rightarrow 0$. At the same time, the single-impurity quality factors $Q^{(1)}=\Omega_\text{Eff}/\Gamma_\text{Eff}$ can be very large, e.g. when identical configuration $\Gamma_\text{Eff}$ is optimized by setting detuning $\delta_\text{LI}^\text{D}(\mathbf{k}\text{=}0)$. In particular, $\Omega_\text{Eff}/\Gamma_\text{Eff} \geq \Omega_\text{I}/\gamma_\text{I}$ for the identical polarization case for a weak, perpendicularly incident drive \cite{SuppInfo}. \crefformat{equation}{Eq.~#2(#1)#3}

\begin{figure}
\begin{center}
\includegraphics[width=	\columnwidth]{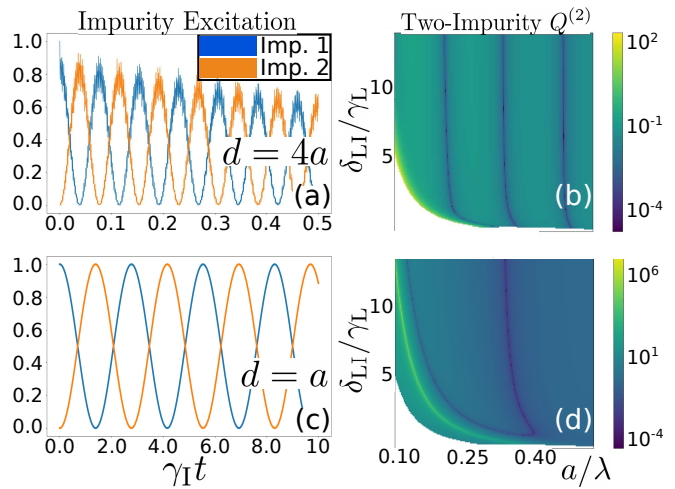}
\caption{The time dependent transfer of excitation between two impurities in arrays with $a=0.1\lambda$ in (a) the orthogonal configuration at distance $d=4a$ (for $d$ dependence, see Fig.\ \ref{fig.Qvsd}) and (c) the identical configuration at $d=a$. (b) Two-impurity quality factor $Q^{(2)}$ as a function of lattice spacing $a$ and detuning $\delta_\text{LI}$ for the orthogonal configuration excitation transfer shown in (a), and (d) $Q^{(2)}$ of the identical configuration shown in (c). In (d), the yellow streak of high $Q^{(2)}$ represents the minimal effective impurity linewidth $\Gamma_\text{Eff}$ predicted by \cref{eq.deltaoptimal} with $\mathbf{k}=0$. Likewise, in (a), $Q^{(2)}$ is maximized for lattice detuning $\delta_\text{LI}$ nearest the band edge. The dark blue bifurcations occur where the free-space and lattice-mediated components of $\Phi_\text{Eff}$ cancel. All arrays are $10\times10$.}
\label{fig.Q}
\end{center}
\end{figure}

We now focus on lattice-mediated interactions between two impurities. When a second impurity is present, the atoms exchange photons via dipole-dipole interactions. This exchange has a lattice-independent component $\phi$, which is simply the free-space dipole-dipole interaction between the impurities \cite{Lehmberg1970}, and a lattice-mediated component, which represents the modification of the dipole-dipole interactions between two impurities due to their interactions with the lattice \cite{SuppInfo}. The impurities thus experience an effective dipole-dipole interaction

\begin{equation}
\Phi_\text{Eff} = \sum_\mathbf{k} \frac{\left(\tilde{J}(\mathbf{k})\text{+}\frac{i}{2}\tilde{\Gamma}(\mathbf{k})\right)\left(\tilde{J}'(\mathbf{k})^*\text{+}\frac{i}{2}\tilde{\Gamma}'(\mathbf{k})^*\right)}{\delta_\text{LI} - J(\mathbf{k}) - \frac{i}{2} \Gamma(\mathbf{k})} + \phi,
\label{eq.Phi}
\end{equation}

\noindent where $\tilde{J}'(\mathbf{k})$ and $\tilde{\Gamma}'(\mathbf{k})$ are the array coupling terms of the second impurity. The quantity $\Phi_\text{Eff}$ is a key metric because it describes the lattice-mediated photon transfer between impurities, analogous to \crefformat{equation}{Eq.~#2(#1)#3}\cref{eq.LMR}, but with the optical driving field replaced by the field of the second impurity. Thus, $\Phi_\text{Eff}$ depends on both the distance between impurities $d$ (dashed line in \crefformat{figure}{Fig.~#2(#1)#3}\cref{fig.Diagram}) and the placement of the impurities within their respective plaquettes. In regimes of large dissipative $\Phi_\text{Eff}$, the system experiences large gain that can be interpreted as parity-time symmetry breaking \cite{Li2019}.

When $\delta_\text{LI}$ is sufficiently above the band edge, photon transfer dynamics between two identical impurities are simply those of two atoms in a Markovian bath, such that their interaction is described by modified excitation transfer rate $\Phi_\text{Eff}$ and decay rate  $\Gamma_\text{Eff}$. This interaction can result in coherent oscillations with large two-impurity quality factors $Q^{(2)} = \mathbf{Re}[\Phi_\text{Eff}]/\Gamma_\text{Eff}$ (Fig.\ \ref{fig.Q}).

\begin{figure}
\begin{center}
\includegraphics[width=	\columnwidth]{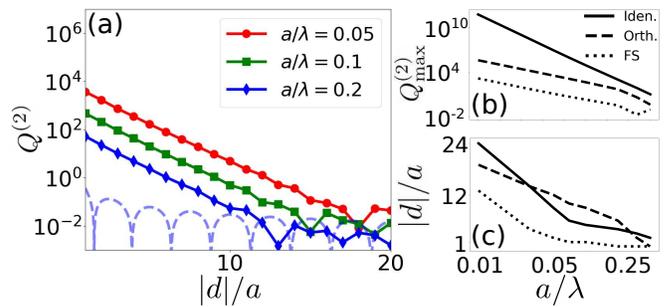}
\caption{(a) Orthogonal configuration two-impurity quality factor $Q^{(2)}$ as a function of distance $d$ between two impurities for various lattice spacings $a$. The value of $Q^{(2)}$ decreases exponentially with $|d|$ due to the photon bound states that mediate this interaction \cite{Douglas2015, Shahmoon2013, Masson2019}. Free-space $Q^{(2)}$ in units of $a/\lambda=0.2$ (dashed light-blue) shown for comparison. (b) $Q^{(2)}_\text{max}\equiv Q^{(2)}(d\text{=}a)$ as a function of $a/\lambda$ for the identical (solid), orthogonal (dashed), and free space (dotted) configurations. (c) The largest number of lattice spacings $|d|/a$ for which a $Q^{(2)}>1$ scales roughly logarithmically with $\lambda/a$.  All arrays are $40\times40$ with detunings $\delta_\text{LI}=1.05 \hspace{0.1cm} \omega_\text{BE}$ and $\delta_\text{LI}=\delta_\text{LI}^\text{D}(\mathbf{k}\text{=}0)$ for the orthogonal and identical configurations, respectively.}
\label{fig.Qvsd}
\end{center}
\end{figure}

\crefformat{figure}{Fig.~#2#1{(a)}#3}\cref{fig.Q} shows an example of such oscillations in the orthogonal configuration with $|d|=0.4\lambda$, $a=0.1\lambda$, and $Q^{(2)} \sim 10^2$. The high frequency, small amplitude modulations are induced by interactions with lattice modes, especially those near the band edge. As this coupling leads to impurity-lattice states outside of our Markovian bath approximation, the analytic value for orthogonal configuration $Q^{(2)}$ in \crefformat{figure}{Figs.~#2#1{(b)}#3}\cref{fig.Q} and \ref{fig.Qvsd} are slight overestimates \cite{Shahmoon2013}, whereas the oscillations of \crefformat{figure}{Fig.~#2#1{(a)}#3}\cref{fig.Q} are exact numerical solutions. \crefformat{figure}{Fig.~#2#1{(b)}#3}\cref{fig.Q} is restricted to $\delta_\text{LI} > 1.05 \hspace{0.1cm} \omega_\text{BE}$ in order to minimize this error. The yellow regions on the left side of the plot show the strong coupling regimes adjacent to the band edge ($Q^{(2)}\sim10^2$), while the dark blue lines represent regions of vanishing $Q^{(2)}$ occurring when the free-space and lattice mediated components of $\Phi_\text{Eff}$ destructively interfere.

2D arrays also facilitate strong coupling between impurities in the identical configuration. The time-dependent, highly-coherent excitation transfer for $\delta_\text{LI}=\delta_\text{LI}^\text{D}(\mathbf{k}=0)$ and $a=0.1\lambda$ is displayed in \crefformat{figure}{Fig.~#2#1{(c)}#3}\cref{fig.Q}. These oscillations feature $Q^{(2)}$ up to $10^5$ and can be described analytically for two impurities in a Markovian bath, as this approximation holds nearly exactly in this regime. In general, identical configuration impurities reach large $Q^{(2)}$ values for $\delta_\text{LI}=\delta_\text{LI}^\text{D}(\mathbf{k}=0)$ (yellow streak in \crefformat{figure}{Fig.~#2#1{(d)}#3}\cref{fig.Q}). This configuration also demonstrates bifurcated regions of low $Q^{(2)}$ due to vanishing $\Phi_\text{Eff}$ (dark blue curve).

The pros and cons of each polarization configuration can be further understood by examining the effect of impurity distance $d$ and lattice spacing $a$ on $Q^{(2)}$. Larger $|d|$ weakens both free-space and array mediated dipole-dipole interactions, reducing $\Phi_\text{Eff}$. In the orthogonal polarization configuration, $Q^{(2)}$ is proportional to $e^{-|d|/\xi}$ for some parameter dependent length scale $\xi$ (\crefformat{figure}{Fig.~#2#1{(a)}#3}\cref{fig.Qvsd}). This scaling is consistent with the width of exponentially localized bound states that the impurities form with array atoms \cite{Douglas2015, Shahmoon2013, Masson2019} and holds until $Q^{(2)}$ approaches its free-space limit (dashed, light-blue curve for $a/\lambda=0.2$), at which point it fluctuates with $\phi$. As discussed above, although the width of these bound states is maximized for $\delta_\text{LI} \approx \omega_\text{BE}$, we set $\delta_\text{LI} = 1.05 \hspace{0.1cm} \omega_\text{BE}$ in order to maintain high impurity excitation and keep the system within the Markovian regime \cite{Shahmoon2013}. The behavior of the identical configuration is similar to that of the orthogonal configuration, but demonstrates larger $Q^{(2)}$ for small $|d|$. However, as the identical configuration relies on the formation of dark states with relatively few array atoms, its $Q^{(2)}$ decreases more rapidly as a function of $|d|$, rendering it the preferable configuration for closely spaced impurities \cite{SuppInfo}.

As this exponential dependence on $|d|$ implies, maximum coupling occurs between impurities in adjacent plaquettes. \crefformat{figure}{Fig.~#2#1{(b)}#3}\cref{fig.Qvsd} displays $Q^{(2)}_\text{max}\equiv Q^{(2)}(d\text{=}a)$. In the identical configuration, $Q^{(2)}_\text{max}$ diverges as $1/a^6$ for small $a$, which is consistent with the $1/a^3$ dipole-dipole interaction strength that mediates both the coupling enhancement and linewidth suppression of the dark state. Similarly, the orthogonal and free-space configurations exhibit a $1/a^3$ scaling, which is consistent with coupling enhancement in a system of relatively static linewidth ($\Gamma_\text{Eff} \approx \gamma_\text{L}$ for $\omega>\omega_\text{BE}$, see \crefformat{figure}{Fig.~#2#1{(b)}#3}\cref{fig.Gamma}). The size of a network of interacting impurity atoms would be limited by the maximum number of lattice spaces $|d|/a$ at which a given $Q^{(2)}$ could be achieved. \crefformat{figure}{Fig.~#2#1{(c)}#3}\cref{fig.Qvsd} shows an approximately logarithmic scaling in $\lambda/a$ for $Q^{(2)}>1$.

Overall, lattice-mediated coupling improves impurity-impurity quality factors $Q^{(2)}$ by several orders of magnitude and extends coupling to tens of lattice sites. While both polarization configurations achieve these effects, we re-emphasize that the identical configuration has greater $Q^{(2)}$ for small $|d|$, while the orthogonal configuration is preferable in the limit of large $|d|$.

In conclusion, we have demonstrated that 2D atom arrays  can effectively mediate between single photons and impurity atoms. We have shown the important role of polarization in the preferential coupling of the impurity to lattice modes. As the optimal detuning for impurity excitation and lattice-mediated two-impurity interactions is above the lattice band edge, the excitation can be localized near the impurity in a controlled way. This allows for impurity-impurity interactions that are substantially stronger and further-reaching than those of free space, resulting in large two-impurity quality factors $Q^{(2)}$ that correspond to coherent two-atom interactions. These results provide a framework 
 for a multilevel atom treatment \cite{Asenjo-Garcia2019}, which can be used to describe coherent switching and quantum gates. These features can be enhanced by the selective excitation of directional, guided lattice modes \cite{Asenjo-Garcia2017}. As the strong coherence and controllable dissipation of this system displays parity-time symmetry breaking \cite{Li2019}, it can also extend to studies of so-called exceptional points \cite{El-Ganainy2018}. Finally, we note that similar effects can be explored in 2D solid state systems corresponding to single- or bi-layer transition metal dichalcogenides \cite{Choi2017}, where excitons with properly engineered band structure can be used to effectively mediate strong interactions between localized impurities. 

\begin{acknowledgments}
We thank Arno Rauschenbeutel for insightful conversations about array-mediated strong interactions with quantum light. This material is based upon work supported by the National Science Foundation Graduate Research Fellowship under Grant No. DGE-1745303. E.S. acknowledges financial support from the Center for New Scientists at the Weizmann Institute of Science. SFY would like to thank the NSF for the CUA PFC grant PHY-1734011 (general work and context), the AFOSR via FA9550-19-1-0233 (detailed calculation of band structures and numerics), and the DOE via DE-SC0020115 (comparison with photonic band structure material).
\end{acknowledgments}

\end{document}


\newcommand{\mean}[1]{\left\langle #1 \right\rangle}

\renewcommand{\thefigure}{S\arabic{figure}}
\renewcommand{\theequation}{S\arabic{equation}}

\title{Supplemental Material: Controlling interactions between quantum emitters using atom arrays}

\author{Taylor L. Patti}
\email[]{taylorpatti@g.harvard.edu}
\affiliation{Department of Physics, Harvard University, Cambridge, Massachusetts 02138, USA}

\author{Dominik S. Wild}
\affiliation{Department of Physics, Harvard University, Cambridge, Massachusetts 02138, USA}

\author{Ephraim Shahmoon}
\affiliation{Department of Physics, Harvard University, Cambridge, Massachusetts 02138, USA}
\affiliation{Department of Chemical \& Biological Physics, Weizmann Institute of Science, Rehovot 761001, Israel}

\author{Mikhail D. Lukin}
\affiliation{Department of Physics, Harvard University, Cambridge, Massachusetts 02138, USA}

\author{Susanne F. Yelin}
\affiliation{Department of Physics, Harvard University, Cambridge, Massachusetts 02138, USA}
\affiliation{Department of Physics, University of Connecticut, Storrs, Connecticut 06269, USA}

\maketitle


\section{Impurity Placement in Array}
\label{sec.Impurity Placement in Array}

We place the impurity atoms interstitially, meaning in the plane of the lattice, but not in a lattice site. Impurity atoms which replace lattice atoms in a lattice site can produce similar effects, but they alter the periodicity of the system and complicate analytic treatment. While the results of this letter focus on impurities at the center of a plaquette, all main text equations besides \crefformat{equation}{Eq.~#2(#1)#3}\cref{eq.deltaoptimal} hold for general, in-plane placement.

\section{Real-Space Atomic Coupling}
\label{sec.Real-Space Atomic Coupling}

The coherent $J(\mathbf{r}_i, \mathbf{r}_j)$ and incoherent $\Gamma(\mathbf{r}_i, \mathbf{r}_j)$  dipole-dipole interaction between two atoms in real-space is given by

\begin{equation}
J(\mathbf{r}_i, \mathbf{r}_j) - \frac{i}{2} \Gamma(\mathbf{r}_i, \mathbf{r}_j) = \text{-}\frac{3 \pi \sqrt{\gamma_i \gamma_j} }{\omega_\text{L}}\hat{d^{\dagger}}_i \cdot\mathbf{G}(\mathbf{r}_i, \mathbf{r}_j,\omega_\text{L}) \cdot \hat{d}_j.
\label{eq.coupling}
\end{equation}

\noindent As discussed in the main text, the mediation of light-atom coupling using atomic arrays has an inherent advantage over solid-state architectures, such as photonic crystals, as it permits us to select an arbitrarily high coherent to incoherent coupling rate $J(\mathbf{r}_i, \mathbf{r}_j)/\Gamma(\mathbf{r}_i, \mathbf{r}_j)$ as atom separation $r = |\mathbf{r}_i - \mathbf{r}_j| \rightarrow 0$. In particular, $\Gamma(\mathbf{r}_i, \mathbf{r}_j) \rightarrow \sqrt{\gamma_i \gamma_j}$ while $J(\mathbf{r}_i, \mathbf{r}_j)$ diverges as $\propto \frac{1}{r^3}$.

We observe this asymptotic behavior by considering free-space Green's function

\begin{equation} 
\label{eq:Greens}
G_{ij}(\mathbf{r}) =
\frac{e^{i \omega r}}{4 \pi r} \left[\left(1 + \frac{i}{\omega r} - \frac{1}{\omega^2 r^2}\right)\delta_{ij} - \left(1 + \frac{3i}{\omega r} - \frac{3}{\omega^2 r^2}\right)\frac{r_i r_j}{r^2}\right] - \frac{\delta(\mathbf{r})}{3 \omega^2} \delta_{ij}.
\end{equation}

\noindent For $r \ll \lambda$, we can take the Taylor expansion of $e^{i \omega r}$ around $r=0$ and keep only the highest order terms, yielding real components $\propto \frac{1}{\omega^2 r^3}$, imaginary diagonal components $\frac{i \omega}{6 \pi}$, and off-diagonal imaginary components of approximately zero.

\section{Impurity-Lattice Coupling}
\label{sec.Impurity-Lattice Coupling}

The dynamics of 2D atomic square lattices and arrays have been discussed in various publications and will not be repeated here. The coupling of a separate impurity atom to these atomic band structures, however, is novel and the topic of this section.

\subsection{$k$-Space Interaction Terms}
\label{sec.$k$-Space Interaction Terms}

The interaction Hamiltonian between an impurity atom with lowering operator $s$, frequency $\omega_\text{I}$, and linewidth $\gamma_\text{I}$, and a collection of lattice atoms with lowering operators $\sigma_i$, frequencies $\omega_\text{L}$, and linewidths $\gamma_\text{L}$ is given generally by

\begin{equation}
H_I^{(1)} = \sum_i \left(J(\mathbf{r}_i, \mathbf{r}_s) - \frac{i}{2}\Gamma(\mathbf{r}_i, \mathbf{r}_s) \right)\sigma^\dagger_i s + \text{c.c.},
\label{eq.H_I}
\end{equation}

\noindent where $J$ and $\Gamma$ are as described by \crefformat{equation}{Eq.~#2(#1)#3}\cref{eq.coupling}. We can transform this relation into $k$-space with the operator substitution $\sigma_i = \sum_\mathbf{k} \sigma_\mathbf{k} e^{i\mathbf{k}\cdot\mathbf{r}_i}$ and identity $\delta(\mathbf{k}) = \sum_i e^{-i\mathbf{k}\cdot \mathbf{r}_i}$, producing equivalent Hamiltonian

\begin{equation}
H_I^{(1)} = \sum_\mathbf{k} \left(\tilde{J}(\mathbf{k}) - \frac{i}{2}\tilde{\Gamma}(\mathbf{k}) \right)\sigma^\dagger_\mathbf{k} s + \text{c.c.},
\label{eq.H_Ik}
\end{equation}

\noindent where $\tilde{J}(\mathbf{k})=\sum_i J(\mathbf{r}_i-\mathbf{r}_s)e^{-\mathbf{k}\cdot\mathbf{r}_i}$ and $\tilde{\Gamma}(\mathbf{k})=\sum_i \Gamma(\mathbf{r}_i-\mathbf{r}_s)e^{-\mathbf{k}\cdot\mathbf{r}_i}$.

For an infinite lattice, we can write this coupling in terms of the momentum-space Green's function according to the relation $\sum_i f(\mathbf{R}_i + \mathbf{\delta})e^{-i\mathbf{k}\cdot (\mathbf{R}_i + \mathbf{\delta})} = \frac{1}{a^2}\sum_i f(\mathbf{k}+\mathbf{G}_i)e^{i\mathbf{G}_i\cdot \mathbf{\delta}}$, where $\mathbf{r}_i=\mathbf{R}_i + \mathbf{\delta}$ for lattice vector $\mathbf{R}_i$ and $\mathbf{G}_i$ are the reciprocal lattice vectors. These relations yield

\begin{subequations}
\begin{eqnarray}
 \frac{\tilde{J}(\mathbf{k})}{\sqrt{\gamma_\text{I}\gamma_\text{L}}} = \frac{-3\pi}{\omega_\text{L}a^2} \hat{d}^\dagger_\text{L} \cdot \sum_i \text{Re}\mathbf{G}(\mathbf{k}\text{+}\mathbf{G}_i)e^{i\mathbf{G}_i\cdot \mathbf{\delta}} \cdot \hat{d}_\text{I},\\
  \frac{\tilde{\Gamma}(\mathbf{k})}{\sqrt{\gamma_\text{I}\gamma_\text{L}}} = \frac{6\pi}{\omega_\text{L}a^2} \hat{d}^\dagger_\text{L} \cdot \sum_i \text{Im}\mathbf{G}(\mathbf{k}\text{+}\mathbf{G}_i)e^{i\mathbf{G}_i\cdot \mathbf{\delta}} \cdot \hat{d}_\text{I}.
\end{eqnarray}
\end{subequations}

\subsection{Modification of Impurity Electromagnetic Environment}
\label{subsec.Modification of Impurity Electromagnetic Environment}

For $\gamma_\text{L} \gg \gamma_\text{I}$, the lattice atoms serve as a Markovian bath for the impurity atoms. Taking the Hamiltonian from \crefformat{equation}{Eq.~#2(#1)#3}\cref{eq.H} and adding both the diagonal terms and an impurity-resonant plane wave driving field which induces weak Rabi frequency $\Omega_\text{L}$ ($\Omega_\text{I}$) on the lattice (impurity) atoms ($\Omega_\text{L}/\gamma_\text{L} \ll 1$, single excitation limit), we find the steady-state solutions of lattice operators $\sigma_\mathbf{k}$ to the resulting Heisenberg-Langevin equations of motion under the harmonic oscillator approximation ($\sigma^\dagger_\mathbf{k}\sigma_\mathbf{k} \approx -1$). Substituting these solutions into the Heisenberg-Langevin equations of motion of the impurity atom, we reduce the problem to a single atom in a modified electromagnetic environment

\begin{equation}
\dot{s} = i\left(\frac{i}{2}\gamma_\text{I} - \Sigma_\text{SE}\right)s - i \Omega_\text{Eff}^* s^\dagger s,
\label{eq.singleatomevolution}
\end{equation}

\noindent where the modifications are a self-energy term (proportional to the square of impurity-lattice interaction) $\Sigma_\text{SE}$ (\crefformat{equation}{Eq.~#2(#1)#3}\cref{eq.SE}) and a lattice mediated field with effective Rabi-frequency (proportional to lattice Rabi drive) $\Omega_\text{Eff}$ (\crefformat{equation}{Eq.~#2(#1)#3}\cref{eq.LMR}).

\subsection{Two-Impurity Interaction}
\label{subsec.Two-Impurity Interaction}

We now derive the dynamics for a system of two impurities with lowering operators $s$ and $q$. Here we assume that the two impurities are of the same species and located at the center of a lattice plaquette, but the formalism follows for impurities of general frequency, linewidth, and placement by following the derivation steps in Sec.\ \ref{sec.$k$-Space Interaction Terms}. The two-impurity interaction Hamiltonian of this symmetric case is

\begin{multline}
H_I^{(2)} = \sum_\mathbf{k} \left(\tilde{J}(\mathbf{k}) - i\frac{\tilde{\Gamma}(\mathbf{k})}{2}\right)\sigma^\dagger_\mathbf{k} s + \sum_\mathbf{k} \left(\tilde{J}'(\mathbf{k}) - i\frac{\tilde{\Gamma}'(\mathbf{k})}{2} \right) \sigma^\dagger_\mathbf{k} q +
\frac{\sqrt{\gamma_\text{s}\gamma_\text{q}}}{\gamma_\text{L}}\left(J(\mathbf{r}_q-\mathbf{r}_s)-i\frac{\Gamma(\mathbf{r}_q-\mathbf{r}_s)}{2}\right)q^\dagger s + \text{c.c.}
\label{eq.twoimpurityH}
\end{multline}

\noindent For the two-impurity interaction of \crefformat{equation}{Eq.~#2(#1)#3}\cref{eq.Phi}, we take the Markov approximation and substitute the steady-states of the impurity atoms as above, however into the equations of motion of the two impurity system.

\section{Additional Notes on Identical Configuration $Q^{(2)}$}
\label{subsec.Identical Q}

\begin{figure}[h]
\begin{center}
\includegraphics[width=	0.3\columnwidth]{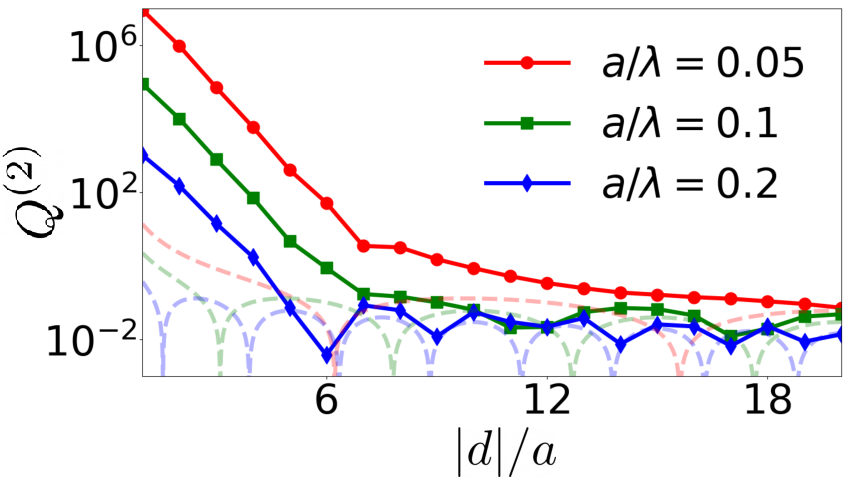}
\caption{Two-impurity quality factor $Q^{(2)}$ vs impurity separation $|d|$ for the identical configuration in a $40\times40$ array with $\delta_\text{LI}=\delta_\text{LI}^\text{D}(\mathbf{k}=0)$ for variouis lattice constants $a$ (solid lines) and for free space impurities (dashed lines, color corresponding to units $a/\lambda$). For small $|d|$, $Q^{(2)}$ is increased by various orders of magnitude over free space. This coupling decreases exponentially in $|d|$ until it either becomes comparable to free-space $Q^{(2)}$, at which point it fluctuates with the two impurity free-space dipole-dipole interactions, or until it surpasses six lattice spacings, at which point the two impurities no longer share adjacent $6\times6$ dark state lattice atoms (see $2\times2$ toy model of Sec.\ \ref{sec.Toy Model}). Details in Sec.\ \ref{subsec.Identical Q}.}
\label{fig.IdenticalPolQvsd}
\end{center}
\end{figure}

The two-impurity quality factor $Q^{(2)}$ of the identical configuration has similar dependence on impurity separation $|d|$ as its orthogonal configuration counterpart (see \crefformat{figure}{Fig.~#2#1{(a)}#3}\cref{fig.Qvsd}), but is multiple orders of magnitude larger for small $|d|$ and extends over fewer lattice sites (Fig.\ \ref{fig.IdenticalPolQvsd}). As the $2\times2$ toy model dark state of Sec.\ \ref{sec.Toy Model} indicates, this is due to strong coupling between the impurities and relatively few nearest neighbor atoms. As identical configuration $Q^{(2)}$ is maximized by driving the impurit dark states, the $Q^{(2)}$ factors of Fig.\ \ref{fig.IdenticalPolQvsd} are tuned to $\delta_\text{LI}=\delta_\text{LI}^\text{D}(\mathbf{k}=0)$ and, as a result, demonstrate exponential drop in $Q^{(2)}$ vs $|d|$ for $|d|/a>6$, after which impurities no longer share array atoms that form non-negligible components of their dark states. The relative compactness of these dark states can be understood via Fig.\ \ref{fig.coupling_vs_size}, which indicates that the electromagnetic environment of an impurity atom in a finite array converges to that of an infinite lattice for arrays $\geq 6\times6$ atoms. When array $Q^{(2)}$ approaches that of free-space coupling, its behavior is dominated by the free-space dipole-dipole interactions $\phi$ between impurities.

\section{$2\times2$ Toy Model and Electromagnetically Induced Transparency}
\label{sec.Toy Model}

\begin{figure}[H]
\begin{center}
\includegraphics[width=	0.4\columnwidth]{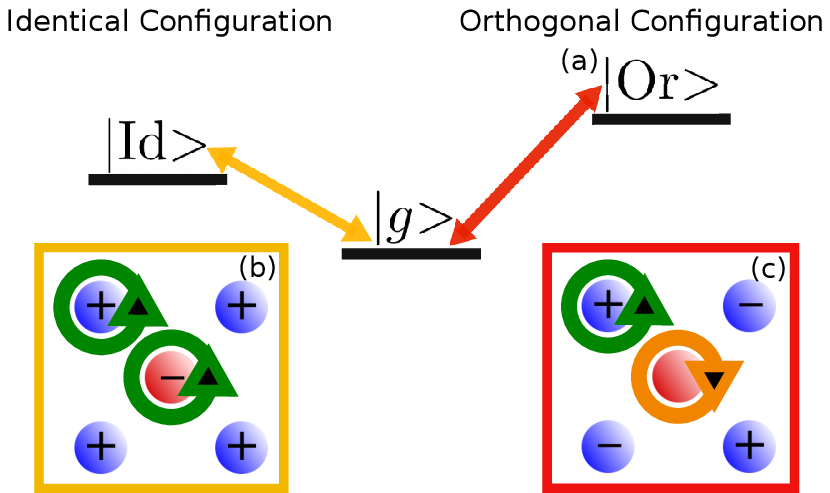}
\caption{(a) An impurity atom in a lattice of right-handed circularly polarized atoms can be driven from ground state $\ket{g}$ with either a right-handed photon (yellow) or a left-handed photon (red), to produce the identical and orthogonal configurations, respectively. (b) $2\times2$ toy model dark $\hat{v}_D$ and (c) bright $\hat{v}_B$ states. Details are given in Sec.\ \ref{sec.Toy Model}}
\label{fig.ToyModel}
\end{center}
\end{figure}

The influence of the array on the impurity atom can be understood by comparing the two limiting cases: the 2x2 array toy model shown in Fig.\ \ref{fig.ToyModel} and the infinite lattice. In the 2x2 lattice (infinite array) case, the impurity with identical circular polarization (yellow arrow \crefformat{figure}{Fig.~#2#1{(a)}#3}\cref{fig.ToyModel}) only couples (couples most strongly) to the mode with lowest energy and highest fluorescence: the in-phase state $\hat{v}_\parallel$ of the 2x2 array (lightcone of the infinite lattice). This forms the dark state $\hat{v}_D$ of \crefformat{figure}{Fig.~#2#1{(b)}#3}\cref{fig.ToyModel}, which we can view as being in the V configuration form of electromagnetically induced transparency (EIT). Likewise, the impurity with orthogonal circular polarization (red arrow \crefformat{figure}{Fig.~#2#1{(a)}#3}\cref{fig.ToyModel}) only couples (couples most strongly) to the mode with highest energy and lowest fluorescence: the out-of-phase state $\hat{v}_\perp$ of the 2x2 array (the band edge of the  infinite lattice). This produces the bright state $\hat{v}_B$ of \crefformat{figure}{Fig.~#2#1{(c)}#3}\cref{fig.ToyModel}. 

\crefformat{figure}{Fig.~#2#1{(b)}#3}

\begin{figure}[h]
\begin{center}
\includegraphics[width=	0.4\columnwidth]{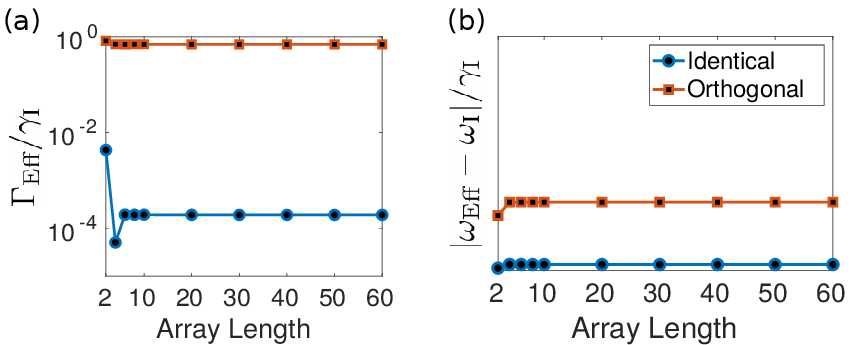}
\caption{Effective impurity linewidth $\Gamma_\text{Eff}/\gamma_\text{I}$ (a) and frequency shift magnitude $|\omega_\text{Eff}-\omega_\text{I}|/\gamma_\text{I}$ (b) in identical (blue circles) and orthogonal (orange squares) circular configurations vs array length with $a=0.2\lambda$. The electromagnetic properties of arrays converge rapidly in lattice size. While the effects of the $2\times2$ toy model are already comparable to those of larger systems, those of the $6\times6$ array are essentially indistinguishable.}
\label{fig.coupling_vs_size}
\end{center}
\end{figure}

We can use the $2\times2$ toy model to understand, both analytically and conceptually, virtually all of the relevant parameters for the modification of the impurity's electromagnetic environment by the array. We start by recognizing that the four lattice atoms of the impurity's immediate plaquette dominate impurity dynamics. For small $a$ such that near field terms dominate and the impurity at the plaquette center, each of the lattice atoms of a $2\times2$ lattice has coherent coupling $\approx 27$ greater than one of the four corner atoms of a $4\times4$ lattice. The electromagnetic environment experienced by the impurity as a function of lattice size is shown in Fig.\ \ref{fig.coupling_vs_size}. Not only is the $2\times2$ model qualitatively similar to larger lattices, arrays as small as $6\times6$ have more or less converged on the effects of arbitrarily large lattices.

\subsection{Subspace of 4 Lattice Atom Modes}
\label{Subspace of 4 Lattice Atom Modes}

We consider a central impurity atom of circular polarization $\nu$ at the center of a $2\times2$ square array of atoms with circular polarization $\mu$ and lattice spacing $a$. The atom order is taken to be counterclockwise such that atom 1 has coordinates $(-a,-a)$, atom 2 $(a,-a)$, atom 3 $(a,a)$, and atom 4 $(-a,a)$. The system is symmetric for the $x$ and $y$-axes and we need only specify the system displacement vectors $\mathbf{r}_1=(a,0)$ or $(0, a)$, $\mathbf{r}_2= (a, a)$, and $\mathbf{r}_3 = (a/2, a/2)$. We can first block diagonalize the Hamiltonian of the array atoms, obtaining eigenmodes:

\begin{equation}
\hat{v}_\parallel = \frac{1}{2}\begin{pmatrix}
    1\\
    1\\
    1\\
    1
\end{pmatrix}, \hspace{0.8cm} \hat{v}_\perp = \frac{1}{2}\begin{pmatrix}
    -1\\
    1\\
    -1\\
    1
\end{pmatrix}, \hspace{0.8cm}
\hat{v}_{M1} = \frac{1}{\sqrt{2}}\begin{pmatrix}
    0\\
    -1\\
    0\\
    1
\end{pmatrix}, \hspace{0.8cm} \hat{v}_{M2} = \frac{1}{\sqrt{2}}\begin{pmatrix}
    -1\\
    0\\
    1\\
    0
\end{pmatrix},
\end{equation}

of eigenvalues:

\begin{subequations}
\begin{align}
\lambda_\parallel &= -\delta_\text{LI} + 2 J_1 + J_2 - \frac{i}{2} \left[\gamma_\text{L} + 2\Gamma_1 + \Gamma_2\right], \\
\lambda_\perp &= - \delta_\text{LI} - 2 J_1 + J_2 - \frac{i}{2} \left[\gamma_\text{L} - 2\Gamma_1 + \Gamma_2\right], \\
\lambda_{M1} &= \lambda_{M2}= -\delta_\text{LI} - J_2 - \frac{i}{2} \left[\gamma_\text{L}-\Gamma_2\right],
\end{align}
\end{subequations}

\noindent where we have simplified $J_{\mu\mu}(\mathbf{r}_1)=J_1$ and $J_{\mu\mu}(\mathbf{r}_2)=J_2$, $\delta_\text{LI} = \omega_\text{I}-\omega_\text{L}$, and other symbol definitions are given in Sec.\ \ref{sec.$k$-Space Interaction Terms}.

We discuss two cases for the lattice $\mu$ and impurity $\nu$ circular polarizations: the identical configuration $\nu=\mu$ and the orthogonal configuration $\nu \perp \mu$. We define $\sqrt{\frac{\gamma_\text{I}}{\gamma_\text{L}}}(J_{\nu\mu}(\mathbf{r}_3),\Gamma_{\nu\mu}(\mathbf{r}_3)) = (J_s, \Gamma_s)$. In the identical configuration, all couplings between the impurity and the 4 lattice atoms are symmetric: $(J_{\nu\mu},\Gamma_{\nu\mu}) = (J_{\mu\nu},\Gamma_{\mu\nu})$. In the orthogonal configuration,  they are anti-symmetic: $(J_{\nu\mu},\Gamma_{\nu\mu}) = (-J_{\mu\nu},-\Gamma_{\mu\nu})$ and $(J_{\mu\nu}(x,-y),\Gamma_{\mu\nu}(x,-y)) = (-J_{\mu\nu}(x,y),-\Gamma_{\mu\nu}(x,y))$.

We develop intuition for the impurity dark state via two perspectives: the first in terms of the steady-state equations of motion for bare states as is used for to analyze coupling to larger arrays in the main text, and the second in terms of a dressed state displaying (EIT). We also briefly discuss the impurity bright state within the first perspective.

\subsection{Perspective 1: Steady-State Equations of Motion}
\label{subsec.Perspective 1: Steady-State Equations of Motion}

\subsubsection{Identical Configuration}
\label{subsubsec.Identical Polarization Block Diagonalization}

If we block diagonalize the interaction for the identical configuration, we find that it  only couples to the in-phase state $\hat{v}_\parallel$. In the weak driving limit, our Hamiltonian reduces to

\begin{equation} \label{eq.identicalH}
H = \begin{bmatrix} -\delta_\text{LI} + J_\parallel - \frac{i}{2}\Gamma_\parallel & \hspace{0.8cm} \tilde{J}_\parallel-\frac{i}{2}\tilde{\Gamma}_\parallel \\
\tilde{J}_\parallel-\frac{i}{2}\tilde{\Gamma}_\parallel & -\frac{i}{2} \gamma_\text{I}
\end{bmatrix}
\end{equation}

\noindent where $J_\parallel= 2J_1+J_2$, $\Gamma_\parallel= \gamma_\text{L}+2\Gamma_1+\Gamma_2$, $\tilde{J}_\parallel=2J_s$, and $\tilde{\Gamma}_\parallel=2\Gamma_s$. We then solve for the self-energy as detailed in Sec.\ \ref{sec.Impurity-Lattice Coupling} and obtain $\Gamma_\text{Eff} = \gamma_\text{I} - 2\mathbf{Im}[\Sigma_\text{SE}]$ where

\begin{equation} \label{eq.identical.SE}
\Sigma_{SE} = \frac{(\tilde{J}_\parallel - \frac{i}{2} \tilde{\Gamma}_\parallel)^2}{\delta_\text{LI}-J_\parallel +\frac{i}{2}\Gamma_\parallel}.
\end{equation}

\noindent Likewise, defining $\Omega_\parallel = 4\times \frac{\Omega_\text{L}}{2}=2\Omega_\text{L}$ for a perpendicular plane wave resonant with the impurity atom, the effective Rabi frequency in the presence of the 2x2 lattice is

\begin{equation} \label{eq.identicalLMR}
\Omega_\text{Eff} = \frac{\left(\tilde{J}_\parallel + \frac{i}{2} \tilde{\Gamma}_\parallel \right)\Omega_\parallel}{\delta_\text{LI}-J_\parallel - \frac{i}{2}\Gamma_\parallel} + \Omega_\text{I}.
\end{equation}

We wish to extremize $\Gamma_\text{Eff}$ with respect to $\delta_\text{LI}$. For the identical configuration

\begin{equation} \label{eq.ImSE}
\mathbf{Im}[\Sigma_{SE}] = -\left[ \frac{(\tilde{J}_\parallel^2-\frac{\tilde{\Gamma}_\parallel^2}{4})\frac{\Gamma_\parallel}{2}+ \tilde{J}_\parallel \tilde{\Gamma}_\parallel(\delta_\text{LI}-J_\parallel)}{(\delta_\text{LI}-J_\parallel)^2 + \frac{\Gamma_\parallel^2}{4}}\right],
\end{equation}

\noindent which, for $\delta_\text{LI} > 0$, prescribes detuning $\delta_\text{LI} = J_\parallel - \tilde{J}_\parallel \Gamma_\parallel/\tilde{\Gamma}_\parallel$ to obtain optimized linewidth

\begin{equation}
\Gamma^\text{Op}_\text{Eff} = \gamma_\text{I} - \frac{\tilde{\Gamma}^2_\parallel}{\Gamma_\parallel}.
\label{eq.GammaEffOp}
\end{equation}

\noindent The refractive effect of this system is then $\mathbf{Re}[\Sigma_\text{SE}] = -\tilde{J}_\parallel\tilde{\Gamma}_\parallel/\Gamma_\parallel$.

As shown previously, the dynamics of impurities in arbitrary arrays approach those of the $2\times2$ toy model for $a/\lambda \ll 1$. In this limit, $\delta_\text{LI}$ scales as $2J_1 + J_2 - 4J_3 \leq -J_3 \rightarrow \infty$. Likewise, in the same limit

\begin{equation}
\Gamma^\text{Op}_\text{Eff} \rightarrow \gamma_\text{I}\left[1 - \frac{4\Gamma^2_3}{\gamma_\text{L}(\gamma_\text{L} + 2\Gamma_1 + \Gamma_2)} \right], \hspace{0.5cm} \Omega^\text{Op}_\text{Eff} \rightarrow \Omega_\text{I} \left[1 - \frac{4\Gamma_3}{\gamma_\text{L} + 2\Gamma_1 + \Gamma_2}\right].
\label{eq.GammaEffOplimit}
\end{equation}

\noindent As all the $\Gamma$ functions approach $\gamma_\text{L}$ in this limit, $\Gamma^\text{Op}_\text{Eff} \rightarrow 0$. $\Omega_\text{Eff}$ also goes to zero by a factor that differs by $\Gamma_3/\gamma_\text{L}$ such that the impurity is decoupled both in terms of decay rate and Rabi drive. As $\Gamma_3/\gamma_\text{L} \leq 1$, with unity holding for $a = 0$, the impurity driving frequency is suppressed less than its decay rate and thus $\Omega_\text{Eff}/\Gamma_\text{Eff} \geq \Omega_\text{I}/\gamma_\text{I}$. Finally, as we would expect, the refractive portion of the self-energy diverges: $\mathbf{Re}[\Sigma^\text{Op}_\text{SE}] \rightarrow -\gamma_\text{I} J_3/\gamma_\text{L} \propto \gamma_\text{I}/r^3$. However, as the group velocity of slowed light is proportional to the two-photon detuning $\delta_\text{LI} = \delta_\text{LI} + \mathbf{Re}[\Sigma_\text{SE}] \approx \delta_\text{LI}$, this effect is actually dominated by the divergence of $\delta_\text{LI}$.

We have identified a limit in which $\Gamma_\text{Eff}, \hspace{0.2cm} \Omega_\text{Eff} \rightarrow 0$ with refractive index $\approx \delta_\text{LI} \rightarrow \infty$, which is clearly the regime of EIT and slow light phenomena. As this regime is limited by small $a$, these effects are lattice spacing limited.

\subsubsection{Orthogonal Configuration}
\label{subsubsec.Orthogonal Polarization Block Diagonalization}

We employ the steps of the preceeding section to the orthogonal configuration eigensystem, obtaining

\begin{equation} \label{eq.oppSE}
\Sigma_{SE} = \frac{ - (\tilde{J}_\perp - \frac{i}{2} \tilde{\Gamma}_\perp)^2}{\delta_\text{LI}-J_\perp +\frac{i}{2}\Gamma_\perp}, \hspace{0.5cm}
\Omega_{\text{Eff}} = \frac{ - \left(\tilde{J}_\perp + \frac{i}{2} \tilde{\Gamma}_\perp \right)\Omega_\perp}{\delta_\text{LI}-J_\perp - \frac{i}{2}\Gamma_\perp},
\end{equation}

\noindent where $J_\perp = \left( - 2J_1 + J_2 \right)$, $\Gamma_\perp = \left(\gamma_\text{L} - 2\Gamma_1 + \Gamma_2 \right)$, $\tilde{J}_\perp = 2J_s$, and $\tilde{\Gamma}_\perp = 2\Gamma_s$. Note that for a perpendicular plane wave, $\Omega_\perp = 2 \times \frac{\Omega_\text{L}}{2} - 2 \times \frac{\Omega_\text{L}}{2} = 0$, and the state decouples from far-field light, fitting with the band edge comparison.

\subsection{Perspective 2: Electromagnetically Induced Transparency Dark State}
\label{subsec.Perspective 2: Electromagnetically Induced Transparency Dark State}

We now wish to find the dressed states of the Hamiltonian of \crefformat{equation}{Eq.~#2(#1)#3}\cref{eq.identicalH}. We solve for the dark and radiant eigenvalues ($\lambda_\text{D}$ and $\lambda_\text{R}$) and unnormalized eigenvectors ($\hat{v}_\text{D}$ and $\hat{v}_\text{R}$) of the system and assuming $\frac{1}{2}\gamma_\text{I} \ll |\delta_\text{LI} - J_\parallel + \frac{i}{2}\Gamma_\parallel|$ we can approximate:

\begin{equation}
\lambda_\text{D} = -\delta_\text{Eff} -\frac{i}{2}\Gamma_\text{Eff}, \hspace{0.5cm} \lambda_\text{R} = -\left(\delta_\text{LI} - J_\parallel + \mathbf{Re}[\Sigma_\text{SE}]\right) - \frac{i}{2}\left(\Gamma_\parallel + 2\mathbf{Im}[\Sigma_\text{SE}]\right),
\label{eq.eigenvalues}
\end{equation}

\noindent where $\lambda_\text{D}$ matches the dark state energy. We also solve for

\begin{equation}
\hat{v}_\text{D} = \begin{pmatrix}
    \tilde{J}_\parallel -\frac{i}{2}\tilde{\Gamma}_\parallel\\
    \delta_\text{LI} - J_\parallel + \frac{i}{2}\Gamma_\parallel
\end{pmatrix}, \hspace{0.4cm} \hat{v}_\text{R} = \begin{pmatrix}
    \delta_\text{LI} - J_\parallel + \frac{i}{2}\Gamma_\parallel\\
    -\left( \tilde{J}_\parallel -\frac{i}{2}\tilde{\Gamma}_\parallel \right)
\end{pmatrix},
\label{eq.eigenvectors}
\end{equation}

\noindent which, for minimized linewidth, can be further simplified to $\hat{v}_\text{D} = (-\alpha, 1)$ and $\hat{v}_\text{R} = (1, \alpha)$, where $\alpha = \frac{\tilde{\Gamma}_\parallel}{\Gamma_\parallel}$.

We introduce a ground state $\ket{g}$ to produce a V system and add two drives: $\Omega_\parallel$ that drives $\ket{g}$ to the bare lattice in-phase mode $\ket{\hat{v}_\parallel}$, and $\Omega_\text{I}$ that drives $\ket{g}$ to the bare impurity mode $\ket{\text{I}}$. The driving terms of our Hamiltonian then become

\begin{equation}
V_\text{drive} = \left(\Omega_\parallel \ket{\hat{v}_\parallel} + \Omega_\text{I}\ket{I}\right)\bra{g} + \text{c.c.}
\label{eq.drive}
\end{equation}

\noindent and our dark state decouples nearly completely, producing net drive $
\Omega_\text{I} - \Omega_\parallel \alpha = \Omega_\text{I} - \Omega_\parallel \frac{\tilde{\Gamma}_\parallel}{\Gamma_\parallel} = \Omega_\text{Eff}$. We observe that $\alpha \rightarrow \frac{1}{2}\sqrt{\frac{\gamma_\text{I}}{\gamma_\text{L}}}$ as $a \rightarrow 0$, at which point $\hat{v}_\parallel$ becomes fully decoupled from the incident light, matching the condition given by the bare state perspective of the previous section. As proven for \crefformat{equation}{Eq.~#2(#1)#3}\cref{eq.GammaEffOp}, $\Gamma_\text{Eff} \rightarrow 0$ at approximately the same rate, illustrating that the dark state becomes decoupled from decay channels as it becomes decoupled from incident far-field light. That is, it approaches a completely dark EIT state under this dressed state formalism. As $\alpha \ll 1$, the dark state is primarily comprised of the impurity bare state with small in-phase lattice mode admixture.

Finally, we briefly examine the radiant state. As $\Sigma_\text{SE} \propto \gamma_\text{I} \ll \gamma_\text{L}$, $\lambda_R$ is approximately equal to the energy/decay of the lattice in-phase mode in the absence of the impurity. Likewise, $\hat{v}_R$ only has small impurity admixture $\propto \alpha$ and $V_\text{drive} \ket{R}$ produces net drive $\Omega_\parallel + \Omega_\text{I} \alpha \approx \Omega_\parallel$. This is fitting with our view of the lattice as a Markovian bath whose dynamics are largely unaffected by those of the impurity.

\bibliographystyle{apsrev4-1}
\bibliography{MyCollection}